\documentclass[aps,pra,twocolumn,groupedaddress,showpacs]{revtex4}

\usepackage{graphicx,color,rotating}
\usepackage{times,txfonts}

\newcommand{\op}[1]{\hat{#1}}
\newcommand{\ket}[1]{  |\,#1 \rangle}
\newcommand{\bra}[1]{ \langle #1\,|}

\newcommand{\ketbra}[2]{\ket{ #1 } \hspace{-1mm} \bra{ #2 } }

\newcommand{\partdd}[1]{ \frac{\partial^2}{\partial #1^2} }

\newcommand{\expect}[1]{\big <  #1  \big>}
\newcommand{\trace}[2]{\textnormal{Tr}_{ #1 } \; #2 \;}
\newcommand{\imag}{{\bf i}}

\begin{document}

\title{Ab-initio phase diagram of ultracold $^{87}$Rb in an one-dimensional two-color superlattice}

\author{Felix Schmitt}
\author{Markus Hild}
\author{Robert Roth}
\affiliation{Institut f\"ur Kernphysik, Technische Universit\"at Darmstadt, 64289 Darmstadt, Germany}

\date{\today}

\begin{abstract}
We investigate the ab-initio phase diagram of ultracold $^{87}$Rb atoms in an one-dimensional two-color superlattice. Using single-particle band structure calculations we map the experimental setup onto the parameters of the Bose-Hubbard model. This ab-initio ansatz allows us to express the phase diagrams in terms of the experimental control parameters, i.e., the intensities of the lasers that form the optical superlattice. In order to solve the many-body problem for experimental system sizes we adopt the density-matrix renormalization-group algorithm. A detailed study of convergence and finite-size effects for all observables is presented. Our results show that all relevant quantum phases, i.e., superfluid, Mott-insulator, and quasi Bose-glass, can be accessed through intensity variation of the lasers alone. However, it turns out that the phase diagram is strongly affected by the longitudinal trapping potential.
\end{abstract}


\maketitle

\section{Introduction}

Ultracold atomic gases in optical lattices have been a topic of active research for about a decade now. One of the research thrusts is the use of these systems as experimental quantum simulators for a variety of lattice models and allow for detailed investigations of strongly correlated quantum systems in a perfectly controllable environment \cite{Zoller,Bloch}. By tuning the laser intensities of the optical lattice alone, one can seamlessly drive a system through quantum phase transitions like the superfluid to Mott-insulator transition \cite{Greiner, Stoeferle}. In so-called two-color superlattices, additional lasers are used to introduce irregular lattice topologies which give rise to exotic quantum phases like the Bose-glass phase \cite{Lye,Fallani}.

\begin{figure}[b]
\includegraphics[width=\columnwidth]{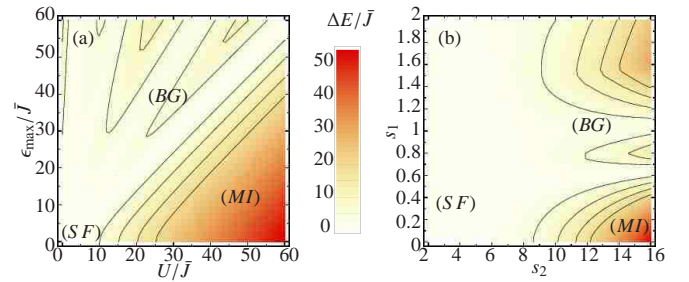}
\caption{(color online) Contour plots of the energy gap computed using DMRG for a commensurate superlattice with $I=N=30$ as (a) a function of the generic Hubbard parameters and (b) a function of the experimental laser intensities. The labels mark the domains of the superfluid (SF) phase, the homogeneous Mott-insulator (MI) phase, and the quasi Bose-glass (BG) phase (taken from Ref.~\cite{Schmitt}).
\label{ComparePic}
}
\end{figure}

Strongly correlated particles in periodic potentials are well described by Hubbard-type models. Together with powerful many-body methods this allows for theoretical studies of the phase diagram of ultracold atomic gases in optical lattices \cite{Roati,Fisher,Scalettar,Rapsch,Damski,RRoth2,RRoth3,RRoth0,RRoth1,RRoth4,Clark,Kollath,Hild,Hild1,Schmitt0,Schmitt,Roux,Roscilde}. However, a one-to-one comparison between experiment and theory has rarely been done so far because these theoretical studies usually adopt the generic parameters of the Hubbard model to span the phase diagram. Such a phase diagram of ultracold bosonic atoms in a two-color superlattice is shown in Fig.~\ref{ComparePic}(a).

In this work we establish a closer link to experiments by computing the phase diagrams with respect to the natural experimental control parameters, which are the intensities ($s_2,s_1$) of the two lasers generating the one-dimensional optical superlattice. Such an experiment-specific phase diagram is shown in Fig.~\ref{ComparePic}(b). In order to predict this type of phase diagram we start with single-particle band structure calculations to extract the Hubbard parameters for a specific experimental setup. Then, the many-body problem is solved using the density-matrix renormalization-group (DMRG) algorithm. In the following, our band structure plus DMRG approach is introduced and benchmarked. We discuss the phase diagram of a specific experimental setup motivated by Refs. \cite{Lye,Fallani} with a focus on its dependence on the transverse trapping frequency $\omega_{\perp}$ and the longitudinal trapping frequency $\omega_{x}$.

\section{1D Bose-Hubbard Model and Band Structure Calculations}

The single-band Bose-Hubbard model \cite{Hubbard} is a widely used framework for studying the ultra-low temperature physics of strongly correlated, neutral atoms in sufficiently deep optical lattices. We assume a one-dimensional lattice with $I$ sites and $N$ bosonic atoms. For each site we define the creation (annihilation) operators $\op{a}^{\dagger}_i$ ($\op{a}^{}_i$) with respect to the localized Wannier states corresponding to the lowest Bloch band. The mean occupation-number at each lattice site is given by $\op{n}_i=\op{a}^{\dagger}_i\op{a}^{}_i$. The Bose-Hubbard Hamiltonian
\begin{eqnarray}\label{HubbardHamil}
	\op{H} =  \sum_{i=1}^{I} && \Big\{ - J_{i,i+1} \; \big( \; \op{a}_{i+1}^{\dagger} \; \op{a}_i^{} + \op{a}_{i}^{\dagger} \; \op{a}^{}_{i+1}\; \big)   \nonumber\\
		                   &&+ \frac{1}{2}  U_{i}  \; \big(\op{n}^{}_i-1 \big) \, \op{n}^{}_i 
		 +  \epsilon_i \; \op{n}^{}_i \Big\}
\end{eqnarray}
accounts for three basic processes: the tunneling of atoms to adjacent sites, the on-site two-body interaction, and the on-site potential energy. The site-dependent Hubbard parameters $J_{i,i+1}$, $U_{i}$, and $\epsilon_i$ define the relative strengths of the individual terms and contain all information about depth and topology of the optical potential, and the interaction between the atoms. The phase diagrams spanned directly by these parameters, typically using $J_{i,i+1}\equiv J$, $U_{i}\equiv U$ and some ansatz for $\epsilon_i$ to account for superlattice structures, are extensively discussed in Refs. \cite{RRoth2,RRoth3,RRoth0,RRoth1,RRoth4,Hild,Hild1,Schmitt0}. More recently, the on-site energies $\epsilon_i$ were calculated directly from the parameters of the optical superlattice to provide a closer connection to experiment  \cite{Roux}.

In this work, our aim is a discussion of the phase diagram using the experimental parameters directly and not the generic Hubbard parameters. To this end, an explicit treatment of the underlying single-particle physics is necessary. Therefore, we start from the optical potential generated by two orthogonal polarized standing-wave laser-fields with wavelengths $\lambda_1$ and $\lambda_2$ and the respective potential depths $s_1$ and $s_2$. Furthermore, we consider an additional harmonic potential with frequency $\omega_x$ accounting for the intensity variation of the optical lattice through the focusing of the laser beams and a magnetic trapping potential. Using the recoil energy $E_{r_i}=\frac{h^2}{2 m \lambda_i^2}$ of atoms with mass $m$ as a natural energy scale and a phase shift $\phi$ between the standing waves, the potential along the x-axes reads:
\begin{eqnarray}\label{OptPot}
 V_{} (x) &=&  s_1  E_{r_1} \sin^2\left(\frac{2 \pi}{\lambda_1} x+ \phi \right) + s_2  E_{r_2} \sin^2\left(\frac{2 \pi}{\lambda_2}x\right)\nonumber\\
 &&+ \frac{1}{2} m \omega_x^2 x^2 \,.
\end{eqnarray}
Throughout this work we consider a setup defined by $\lambda_2=800$ nm and $s_2$ for the primary laser generating the optical lattice potential and $\lambda_1=1000$ nm and $s_1$ for the secondary laser generating the two-color superlattice topology with a phase shift of $\phi=\pi/4$.This leads to the commensurate superlattice that was also used in previous publications \cite{RRoth0,RRoth1,Hild,Hild1,Schmitt0,Schmitt}.

Before we are able to extract the Hubbard parameters for a given potential, we have to determine the localized Wannier functions via a single-particle band structure calculation. For a periodic potential ($s_1=0$, $\omega_{x}=0$) with $I$ sites, we numerically obtain the solutions for the Bloch functions $\psi_k(x)$ in the lowest energy band. The quasimomenta are quantized with respect to the size $I$ of the optical lattice and are labeled by $k$ with $k=0,1,\cdots, I-1$. A Fourier transformation of the Bloch functions with respect to the quasimomenta in the subspace of the lowest energy band leads to the Wannier functions
\begin{equation}\label{wannier}
 w_{i}(x)= \frac{1}{\sqrt{I}} \sum_{k=0}^{I-1} \psi_{k}(x) e^{-\imag \frac{2 \pi}{I} k i} e^{\imag \varphi_{k}} \,.
\end{equation}
The arbitrary phases $\varphi_{k}$ are chosen such that the resulting Wannier functions are maximally localized at their individual lattice site $i$.
Using these maximally localized Wannier functions, the Hubbard parameters are obtained via the matrix elements of the individual terms of the real-space Hamiltonian \cite{Jaksch}
\begin{eqnarray}
 -J_{i,j} &=& \int dx \;  w_{i}^{\ast}(x) \left( -\frac{\hbar^2}{2m} \partdd{x} + V_{}(x) \right)  w_{{j}}(x)\nonumber\\
 \epsilon_{i}^{}  &=& \int dx \;  w_{i}^{\ast}(x) \left( -\frac{\hbar^2}{2m} \partdd{x} + V_{}(x) \right)  w_{i}(x)\label{hubbardparam}\\
 U_{i}  &=& 2 \, \omega_{\bot} \hbar\, a_s\,  \int dx \; |w_{i}(x)|^4 \nonumber \,.
\end{eqnarray}
The contact interaction is defined by the three-dimensional s-wave scattering length $a_s$. The transverse directions are integrated out assuming Gaussian wavefunctions with frequencies $\omega_y=\omega_z=\omega_{\bot}$ \cite{Bloch}. This one-dimensional description is valid as long as the tunneling in the transverse directions is strongly suppressed, i.e., as long as the laser intensities in these directions are sufficiently large.  Following Refs. \cite{Lye,Fallani} we consider a gas of $^{87}$Rb atoms with s-wave scattering length $a_s = 109 \; r_{\text{Bohr}}$ and we assume a transverse trapping frequency $\omega_{\perp}=2 \pi \times 17 $ kHz. In the last part of this work we will discuss the changes in the phase diagram induced by a different value of $\omega_{\perp}$.

As a fir$I$st application of our band structure calculations we validate the single-band approximation in the Hubbard model. As a by-product from the calculations of the Hubbard parameters we obtain from the single-particle band structure calculation the energy gap $\Delta \mathcal{E}$ between the first and the second band of Bloch functions. In Table \ref{tab0} we list some values of $\Delta \mathcal{E}$ together with the Hubbard parameters $U$ and $J$ in the relevant parameter range. Since $\Delta \mathcal{E}$ is always about one order of magnitude larger than $U$ and $J$, we conclude that excitations to energetically higher Bloch bands induced by tunneling or interaction can be neglected even for shallow optical lattices. 

\begin{table}
\begin{tabular}{|c|c|c|c|c|c|c|c|c|}
\hline
$s_2$										&	$2$		&	$4$		&	$6$		&	$8$		&	$10$		&	$12$		&	$14$	&	$16$	\\
\hline
$\Delta \mathcal{E} / E_{r_2}$	&	$1.00$	&	$1.97$	&	$2.90$	&	$3.77$	&	$4.57$		&	$5.31$		&	$5.97$	&	$6.58$\\
\hline
$U / E_{r_2}$							&	$0.143$&	$0.205$&	$0.237$&	$0.261$&	$0.280$	&	$0.297$	&	$0.311$&	$0.323$	\\
\hline
$J / E_{r_2}$								&	$0.143$&$0.086$&	$0.051$&$0.031$&	$0.019$	&	$0.012$	&	$0.008$&	$0.005$	\\
\hline
$U / J$										&	$1.138$&	$2.407$&	$4.671$&	$8.480$&	$14.62$	&	$24.22$	&	$38.84$&	$60.64$	\\
\hline
\end{tabular}
\caption{\label{tab0} Calculations of the Hubbard parameters $U$ and $J$ and the energy gap $\Delta \mathcal{E}$ between first and second Bloch band for a homogeneous lattice ($s_1=0$, $\omega_x=0$) with $\lambda_2=800$ nm, $\omega_{\perp}=2 \pi \times 17$ kHz, $a_s=109 \; r_{\text{Bohr}}$, and mass $m$ of $^{87}$Rb .}
\end{table}

As a second application of our band structure calculations we check for the validity of the restriction to nearest-neighbor tunneling and on-site two-body interactions. For the weakly and the strongly interacting regime, we calculate the respective matrix elements of the Hubbard Hamiltonian using Eqs.~(\ref{hubbardparam}) and $U_{i,j}  = 2 \, \omega_{\bot} \hbar\, a_s\,  \int dx \; |w_{i}(x)|^2 |w_{j}(x)|^2$ for the interaction term. The results are shown in Table \ref{tab1}. Even in the weakly interacting regime ($s_2=2$), the nearest-neighbor tunneling exceeds more-distant tunneling processes by at least one order of magnitude. The interaction matrix element for neighboring lattice sites is already two orders of magnitude smaller than the on-site interaction matrix element. In the strongly interacting regime ($s_2=10$) we already have two orders of magnitude between $J_{i,i+1}$ and $J_{i,i+2}$ and five orders of magnitude between $U_{i,i}$ and $U_{i,i+1}$.  Since we focus on the intermediate and strong interaction regime, the restriction to $J_{i,i+1}$ and $U_i$ is well justified.

\begin{table}
\begin{tabular}{|c|c|c|c|c|c|c|}
\hline
\multicolumn{7}{|c|}{$s_2=2$  ($U_{i,i}/J_{i,i+1}=1.38$)}\\
\hline
 $|i-j|$	&	$0$	&	$1$	&	$2$	 &	$3$	  &	$4$	&	$5$	\\
\hline
$J_{i,j}/E_{r_2}$&	-	&	$0.1428$	&	$-0.02$	 &$0.0048$	  &   $-0.0014 $&$\approx10^{-4}$\\
\hline
$U_{i,j}/E_{r_2}$&	$0.162$	&    $0.0035$	&$\approx10^{-4}$&$\approx10^{-5}$&$\approx10^{-6}$ &$\approx10^{-7}$\\
\hline\hline
\multicolumn{7}{|c|}{$s_2=10$  ($U_{i,i}/J_{i,i+1}=14.62$)}\\
\hline
 $|i-j|$	&	$0$	&	$1$	 &	$2$	  &	$3$	  &	$4$	&	$5$	\\
\hline
 $J_{i,j}/E_{r_2}$&	-	&    $0.0192$	 &$\approx-10^{-4}$&$\approx10^{-6}$&$\approx-10^{-8}$&$\approx10^{-9}$\\
\hline
 $U_{i,j}/E_{r_2}$&	$0.28$	&$\approx10^{-5}$&$\approx10^{-8}$&$<10^{-12}$&$<10^{-12}$ &$<10^{-12}$\\
\hline
\end{tabular}
\caption{\label{tab1}Higher order tunneling and interaction energies for a homogeneous lattice ($s_1=0$, $\omega_x=0$) with $\lambda_2=800$ nm,  $\omega_{\perp}=2 \pi \times 17$ kHz, $a_s=109 \; r_{\text{Bohr}}$, and mass $m$ of $^{87}$Rb .}
\end{table}

So far we have discussed the limit of a homogeneous optical lattice. As soon as the secondary laser which generates the superlattice, or an additional harmonic potential are taken into account, a straight-forward band structure calculation is no longer doable, because Bloch functions are only defined for strictly periodic potentials. Therefore, in order to extract site-dependent Hubbard parameters also for an inhomogeneous lattice we are limited to an approximate scheme to obtain localized Wannier functions. We use two different approaches to extract the site-dependent Hubbard parameters.

\begin{figure}
\includegraphics[width=1\columnwidth]{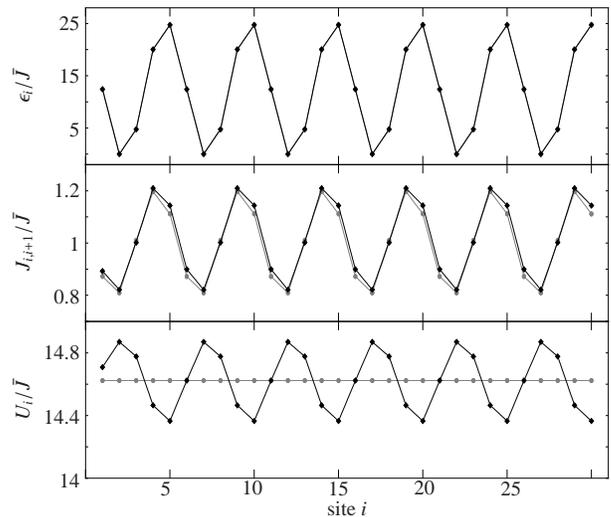}
\caption{\label{FIGHubbardParameters2D}Site-dependent Hubbard parameters obtained from band structure calculations for a two-color superlattice. Simple ansatz  (gray symbols) and calculations with individual Wannier functions (black symbols), both for $s_2=10$ and $s_1=1$. Lines to guide the eye.}
\end{figure}

As a simple ansatz, we consider the secondary laser as a perturbation of the strong primary laser ($s_1 \ll s_2$). The Wannier functions are extracted from a conventional band structure calculation for a homogeneous lattice defined by the primary laser alone. In this approximation the Wannier functions are identical for each lattice site. Using these Wannier functions the Hubbard parameters of each site of the superlattice are computed. The site-dependence of the parameters thus results exclusively from the superlattice potential $V_{}(x)$ entering into the matrix elements (\ref{hubbardparam}) and not from a site-dependence of the Wannier functions themselves. As a result, the parameter $U_i$ characterizing the on-site interaction remains constant for all lattice sites. An exemplary set of site-dependent Hubbard parameters calculated in this scheme is shown in Fig.~\ref{FIGHubbardParameters2D}. Please note that we always subtract a global energy constant from the Hamiltonian to set $\epsilon_{{\text{min}}}= \text{min} \{ \epsilon_i \} =0$.

In a more sophisticated scheme we determine the site-dependent Wannier functions individually for each site of the inhomogeneous lattice using a standard band structure calculation for a periodic lattice with a lattice amplitude defined by the local depth of the inhomogeneous potential at that particular site. In this way, the shape of the Wannier functions depends nontrivially on the local structure of the superlattice potential. The only reason why the set of Wannier functions determined in this way cannot be considered as an exact set of localized basis functions results from the minimal violation of orthogonality for the Wannier functions of neighboring sites. Their mutual overlap is nonzero but always below $1 \%$ in the parameter regime considered in all our calculations. Using these individual localized Wannier functions all site-dependent Hubbard parameters are computed without further approximations. An exemplary set of results is also shown in Fig.~\ref{FIGHubbardParameters2D}. 

The comparison of the site-dependent Hubbard parameters $\epsilon_i$ and $J_{i,i+1}$ obtained by the two schemes shows very little difference. This leads to the conclusion that the second scheme provides a sufficiently accurate description of the Hubbard parameters in the parameter range under consideration, simply because the change induced by first, much cruder approximation is small.

The dominant effect on the Hubbard parameters induced by the superlattice is the spatial variation of the on-site energies $\epsilon_{i}$. This is in agreement with the approximation of the superlattice through this parameter alone \cite{RRoth0,RRoth1,Hild,Hild1,Schmitt0}. However, also the tunneling matrix element $J_{i,i+1}$, essentially depending on the height of the potential barrier between sites $i$ and $i+1$, varies significantly. The on-site interaction matrix element $U_{i}$ exhibits only a weak variation which is introduced by the site-dependence of the Wannier functions in our second scheme.

\section{\label{secDMRG} Density-Matrix Renormalization-Group}

\begin{figure}
\includegraphics[width=1\columnwidth]{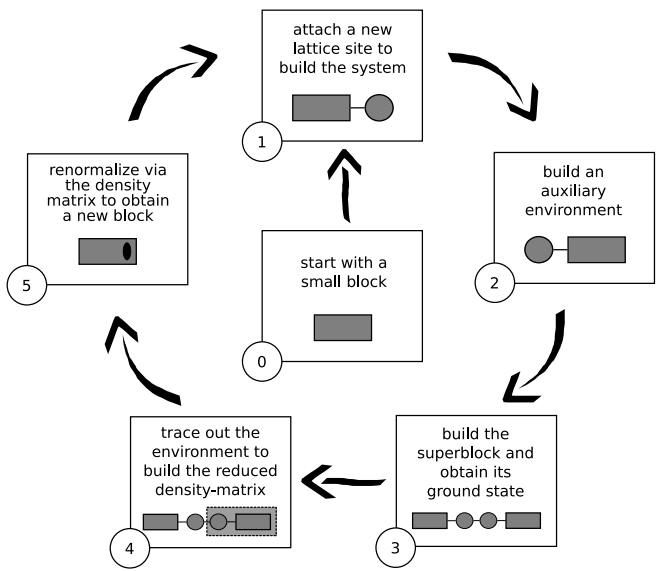}
\caption{
\label{DMRGcycle} Sketch of the DMRG cycle. The black dot in step 5 marks the additional site that was attached without increasing the dimension of the Hilbert space of the block. For details see text.
}
\end{figure}

We solve the many-body problem associated with the Bose-Hubbard Hamiltonian via the density-matrix renormalization-group (DMRG) algorithm \cite{White,Schollwoeck} which is among the most powerful quasi-exact methods available for one-dimensional lattice models. The so-called infinite-size DMRG algorithm is based on an iterative growing procedure. The algorithm is schematically depicted in Fig.~\ref{DMRGcycle}. The individual steps are:
(0) We start with a block composed of $I_{{\text b}}$ sites and up to $N_{\text{b}}$ particles described in a Fock space $\mathcal F_{\text{b}}$ of dimension $D_{{\text b}}$. Since the Hubbard Hamiltonian conserves the particle number, the matrix representation of the block Hamiltonian has a block-diagonal form. Each block of the matrix corresponds to a Hilbert space with a fixed particle number.
(1) To the block we attach an additional lattice site with up to $N_{\text{s}}$ particles described in a Fock space $\mathcal F_{\text{s}}$ to build the Fock space of the system $\mathcal{F}_{\text{sys}}=\mathcal{F}_{{\text b}}\otimes\mathcal{F}_{{\text s}}$ with dimension $D_{\text{sys}}=D_{\text{b}} D_{{\text s}}$.
Again the matrix representation of the system Hamiltonian is block diagonal.
(2+3) In order to simulate a larger lattice, the system is coupled to an analogously constructed environment yielding the superblock $\mathcal{H}_{\text{super}}=\mathcal{F}_{\text{sys}}\otimes \mathcal{F}_{\text{env}}$ of dimension $D_{\text{super}}$ which is projected to a fixed total particle number, satisfying $N/I=1$ in our case.
(3) The ground state $\ket{\psi_0}$ is obtained by diagonalizing the superblock Hamiltonian where one can exploit the sparseness of the Hamilton matrix and use efficient Lanzcos or Jacobi-Davidson algorithms.
(4) The reduced density-matrix is formed by tracing out the environment $\op{\rho}^{\text{red}}=\trace{\text{env}}{\ketbra{\psi_0}{\psi_0}}$. 
(4+5) The $D_{{\text b}}$ eigenvectors of the reduced density-matrix for the largest eigenvalues are used to span the Fock space for a new block $\tilde{\mathcal{F}}_{{\text b}}$ of length $\tilde{I}_{{\text b}} =I_{{\text b}}+1$. These eigenvectors build a non-unitary transformation matrix $\mathcal{O}$ which is employed to construct the new block Hamiltonian $\tilde{H}_{{\text b}}=\mathcal{O}^{\dagger} H_{{\text sys}} \mathcal{O}$. All operators coupling the system to the environment (which will later couple the new block to the new site) and all observables have to be transformed accordingly. This cycle is repeated until the final length of the lattice is reached.

The key feature of this algorithm lies in the use of the eigenvectors with largest eigenvalues of the reduced density-matrix as a new, truncated basis for the new block. One can show that this procedure yields an optimized wavefunction, gives the best approximation to expectation values of observables, and preserves a maximum of entanglement between system end environment \cite{Schollwoeck}.

The error in the DMRG algorithm is due to the loss of information during the non-unitary basis transformations. It can be estimated by summing up the eigenvalues of the discarded eigenvectors. A smaller sum consequently means a smaller loss of information. In addition one has to consider the restriction to a maximal number of particles max$\{n_i\}$ per lattice site. In the complete Hilbert space this would be equal the total number of particles $N$. In general, the stronger the correlations between the particles, i.e., the larger $\bar{U}$, the smaller the sum of the residual eigenvalues and the better the approximation.

If disorder is introduced, then only at the very last step of the growing procedure the full information about the superlattice topology is available to the Hamiltonian. This leads to a poor approximation of the ground state when using the infinite-size algorithm only. As an improvement the finite-size DMRG is applied. After a complete run of the infinite-size algorithm up to the desired length of the lattice, the length of the superblock is kept fixed and the system grows on the expense of the environment and vice versa. During a back and forth sweeping, the superlattice topology is sampled while the Hamiltonian always takes the whole lattice into account. The sweeping continues until all observables are converged.

The way the transformation matrices $\mathcal O$ are constructed is not uniquely defined by the DMRG algorithm. We would like to emphasize that the reduced density-matrix $\rho^{\text{red}}$ is block diagonal and each block has a well defined particle number. One can either use the eigenvectors with the largest eigenvalues for each subspace of $\rho^{\text{red}}$, or one can strictly use the first $D_{{\text b}}$ eigenvectors with the largest eigenvalues not accounting for the block-diagonal structure of $\rho^{\text{red}}$. In the first scheme one might discard eigenvectors with sizable eigenvalues if the respective subspace has reached its preassigned dimension. In the second scheme one might discard a complete subspace of a certain particle number in case it has no eigenvector with corresponding eigenvalue among the largest $D_{{\text b}}$ eigenvalues. If in a subsequent step of the finite size algorithm this subspace becomes important again, this might prevent the algorithm from converging to the proper ground state. This can be overcome by adding noise to the transformation matrices $\mathcal O$ during the first few sweeps in the finite-size algorithm with the aim of recovering lost subspaces again \cite{Garnet}. As a third strategy one can keep at least one or a few eigenstates from each subspace even if their eigenvalues are not among the largest $D_{\text b}$ eigenvalues. We employ the first strategy because it is technically very convenient. However, we checked individual eigenspectra of $\rho^{\text{red}}$ and conformed that none of the discarded eigenvectors had sizable eigenvalues.

\section{Observables}

In this section, we introduce the set of observables we employ to distinguish the different quantum phases.

\paragraph*{Maximum Number Fluctuation.}
The number fluctuation at lattice site $i$ is given by the variance of the occupation number
\begin{equation}
\sigma_i = \sqrt{\expect{\op{n}_i^2} - \expect{\op{n}_i}^2 } \, .
\end{equation}
The number fluctuation provides information about the local mobility of the atoms in the optical lattice. In order to reduce the amount of information we only consider the maximum number fluctuation through the lattice
\begin{equation}
\sigma_{\text{max}} = \text{max}\{ \sigma_i \} \,.
\end{equation}

\paragraph*{Condensate Fraction.}
In order to determine the fraction of atoms that undergo Bose-Einstein condensation we adopt the Onsager-Penrose criterion \cite{Penrose} and calculate the natural orbitals via the eigensystem of the one-body density-matrix $\rho^{(1)}_{ij}=\expect{\op{a}^{\dagger}_i \op{a}_j}$. The largest eigenvalue $N_c$ of the one-body density-matrix is associated with the number of condensed atoms and defines the condensate fraction
\begin{equation}
f_c=\frac{N_c}{N} \,.
\end{equation}

\paragraph*{Visibility.}
In the experiment, most information about the atoms in the optical lattice is extracted from the interference pattern obtained by the time-of-flight method. The interference pattern $\mathcal I (\delta)$ is intimately connected to the quasi-momentum structure of the many-body state and can be calculated from the Fourier transformation of the one-body density-matrix \cite{RRoth2}
\begin{equation}
\mathcal I (\delta) = \frac{1}{I} \sum_{i,j=1}^I e^{\imag (i-j)\delta} \rho^{(1)}_{ij} \,.
\end{equation}
The visibility of the interference fringes $\nu$ is obtained from the maxima and minima of the interference pattern
\begin{equation}
\nu = \frac{\text{max} \{ \mathcal I (\delta) \} - \text{min} \{ \mathcal I (\delta) \} }{\text{max} \{ \mathcal I (\delta) \} + \text{min} \{ \mathcal I (\delta) \} } \,.
\end{equation}

\paragraph*{Energy Gap.}
Measuring the excitation spectrum of the system also provides a sensitive probe for the different quantum phases. In the experiment one employs two-photon Bragg spectroscopy via an intensity modulation of the optical lattice. The width of the central interference peak is used as a measure of the energy transfer into the atomic cloud \cite{Stoeferle}. The detailed structure of the excitation spectrum has been investigated theoretically \cite{Hild,Hild1,Clark,Kollath}. Basic information about the excitation spectrum is given by the energy gap $\Delta E$, which is the minimum amount of energy needed to excite the system. It is defined by the difference between the energy of the first excited state and the ground state
\begin{equation}
 \Delta E = E_1 - E_0 \,.
\end{equation}

An inherent complication in the DMRG framework is the calculation of observables. This is because a DMRG calculation does not yield eigenstates of the Hamiltonian in a simple occupation-number basis representation which could be used to compute observables directly.  Rather, the matrix representations of all observables have to be dragged through all the cycles of the DRMG algorithm, i.e.,  they have to undergo all the lossy non-unitary basis rotations. This is why we thoroughly test our DMRG calculations for convergence.

\section{Benchmark of the DMRG algorithm}

\subsection{Convergence}

\begin{figure*}
\includegraphics[width=1\textwidth]{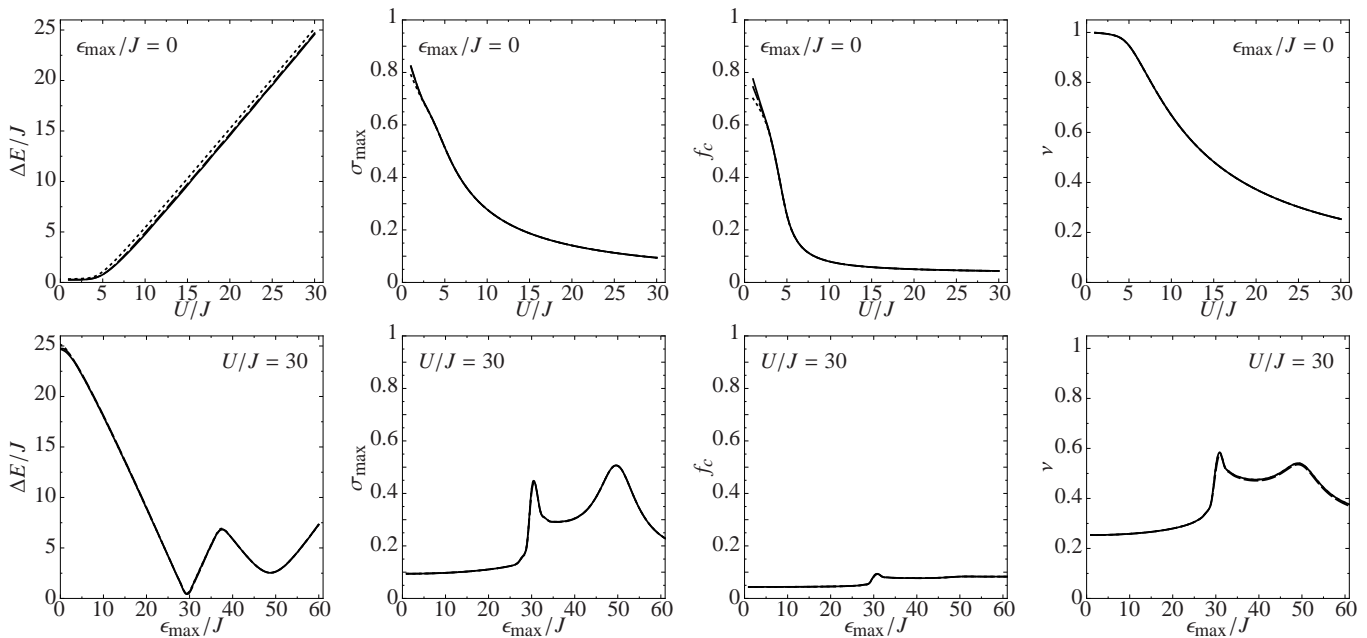}
\caption{Benchmark of the convergence of the DMRG calculations. From left to right: energy gap $\Delta E$, maximum number fluctuation $\sigma_{\text{max}}$, condensate fraction $f_c$, and visibility $\nu$ for a  $I=N=30$ lattice. Upper panels: SF-MI transition at fixed $\epsilon_{\text{max}}/J=0$. Lower panels: MI-BG transition at fixed $U/J=30$. All plots show three lines corresponding to a different basis choice: DMRG-A (dotted), DMRG-B (dashed), and DMRG-C (solid).
\label{DMRGconvergence}
}
\end{figure*}

Before we employ the DMRG algorithm to compute phase diagrams for realistic lattice sizes and particle numbers, we have to assess the precision of the numerical DMRG results. We follow a twofold strategy.

First, we compare results for the various observables obtained by DMRG calculations with results from an exact diagonalization scheme \cite{RRoth0,RRoth1} for a small system with $I=N=10$, where the latter calculations are feasible. In the phase diagram shown in Fig.~\ref{ComparePic}(a) we observed an error of the DMRG calculation below $1\%$ for all observables at $U/J>3$ already for a small DMRG basis with dimension $D_{\text{super}}=338$. The complete Hilbert space used in the exact diagonalization scheme has a dimension of $D=92378$ for the $I=N=10$ system.

Second, in order to validate the results of our DMRG calculations for larger lattices, where no exact calculations in the complete Hilbert space are available, we study the dependence of the DMRG results on basis sizes and particle number truncations used in the algorithm \cite{White,Schollwoeck}. If the results for all observables do not change while the bases size is increased further, the calculation is converged to the exact result. The different basis sets we employ are summarized in Table \ref{tab2}, where max$\{n_i\}$ is the maximum number of particles per lattice site included in the basis.

\begin{table}
 \begin{tabular}{|c|c|c|c|c|}
\hline
	&	$D_{\text{b}}$	&	$D_{\text{s}}$  & $D_{\text{super}}$ & max$\{n_i\}$\\
\hline
DMRG-A	&	21	&	126	&	446		&	5\\
\hline
DMRG-B	&	56	&	336	&	5073	&	5\\
\hline
DMRG-C	&	210	&	1470&	68356	& 6\\
\hline
\end{tabular}
\caption{\label{tab2} Different bases used for studying the convergence of the DRMG calculations. See text for details.}
\end{table}

For all following calculations we applied three sweeps in the finite-size algorithm. For simplicity we consider straight lines through the parameter plane shown in Fig.  \ref{ComparePic}(a).

\paragraph*{Superfluid to Mott-insulator ($\epsilon_{\text{max}}=0$).}
The upper row of images in Fig.~\ref{DMRGconvergence} shows the observables across the superfluid to Mott-insulator phase transition calculated using the three bases specified in Tab. \ref{tab2}. Since the DMRG algorithm is tailored to describe strongly correlated systems, we expect better agreement of the three different calculations with increasing $U/J$. Apart from the energy gap this is confirmed by our calculations. Only for $U/J \lesssim 3$ we observe small differences between the calculations for $\sigma_{\text{max}}$ and $f_c$. For all values of $U/J$ the energy gap is slightly larger when employing the DMRG-A basis. This is because we do not explicitly target at the first excited state for the calculation of the energy gap. Although the ground state has already converged even for the small DMRG-A basis, the first excited state needs a larger basis to converge as well.

\paragraph*{Mott-insulator to quasi Bose-glass ($U/J=30$).}
We already pointed out the importance to use the finite-size DMRG algorithm in order to obtain a converged ground state especially when irregularities in the optical lattice are considered. The results of the observables through the Mott-insulator to quasi Bose-glass transition depicted in the lower panel of Fig.~\ref{DMRGconvergence} show that the finite-size algorithm is perfectly converged for all values of $\epsilon_{\text{max}}/J$ already for the DMRG-A basis.

Since, in this manuscript the focus is on the regime of intermediate and strong interactions, we conclude from our findings that already the DMRG-A basis is suitable to approximate all observables with sufficient precision. Nevertheless, we decided to use the larger DMRG-B basis for all following calculations. Calculations based on this basis are still numerically feasible on a desktop PC while providing good results also in the weakly interacting limit.

\subsection{Finite-Size Analysis}

\begin{figure*}
\includegraphics[width=1\textwidth]{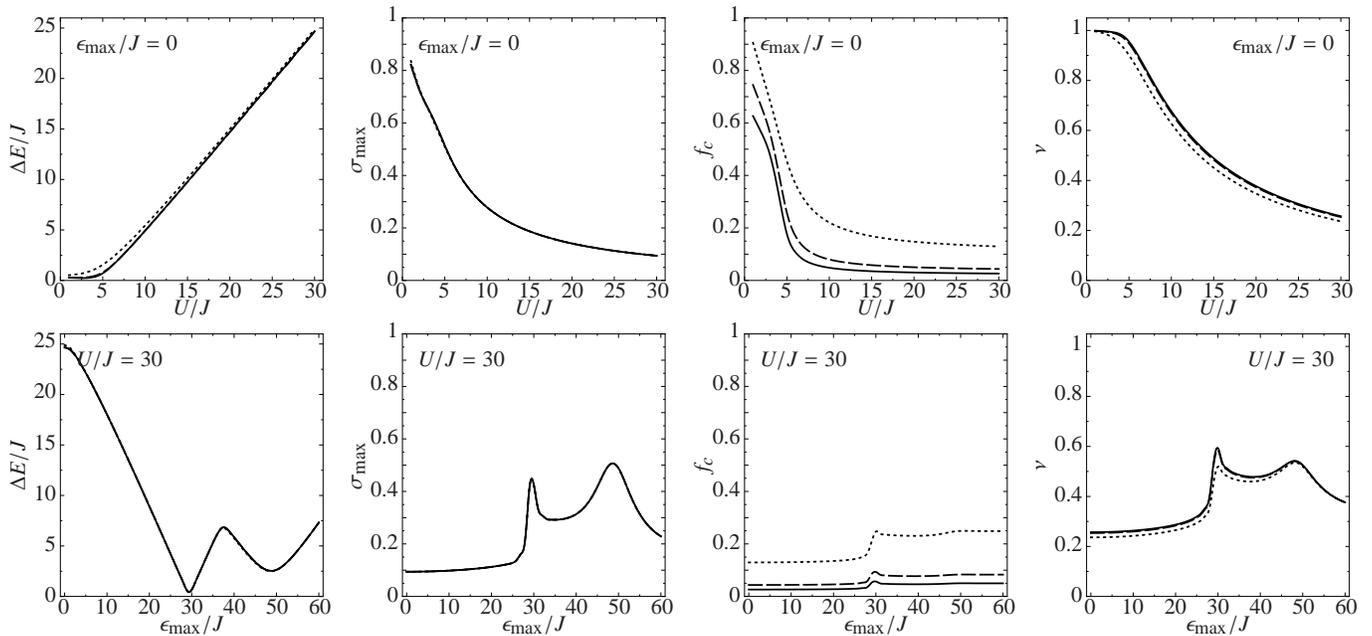}
\caption{\label{DMRGscaling}Finite size analysis for $I=N=10$ (dotted), $I=N=30$ (dashed), and $I=N=50$ (solid). All calculated with the DMRG-B basis. From left to right: energy gap $\Delta E$, maximum number fluctuation $\sigma_{\text{max}}$, condensate fraction $f_c$, and visibility $\nu$. Upper panels: SF-MI transition at fixed $\epsilon_{\text{max}}/J=0$. Lower panels: MI-BG transition at fixed $U/J=30$. 
}
\end{figure*}

We also have to address the dependence of the observables observables on the size of the system. Current experiments typically have between $1.5 \cdot 10^4$ to $2 \cdot 10^5$ atoms in the optical lattices \cite{Lye,Fallani}, i.e., roughly between 25-60 atoms in each one-dimensional array of sites. Thus, we will compare DMRG calculations for $I=N=10$, $I=N=30$, and $I=N=60$ all using the DMRG-B basis. In analogy to the above discussion, we show plots through the superfluid to Mott-insulator transition and the Mott-insulator to quasi Bose-glass transition.

\paragraph{Superfluid to Mott-insulator ($\epsilon_{\text{max}}/J=0$).}
The results are shown in the upper row of Fig.~\ref{DMRGscaling}. By definition, the maximum number fluctuation is a local observable which is calculated at one individual lattice site and is, therefore, practically independent of the size of the lattice. The energy gap as well as the visibility show only small differences between the small and the two larger lattices indicating a minor dependence on length of the lattice for those observables. However, the condensate fraction depends systematically on the size of the lattice. The larger the lattice is, the steeper is the decrease of $f_c$ around $U/J\approx 5$. One can easily show that for $U/J \rightarrow \infty$ and $I=N$ the condensate fraction scales like $f_c \propto 1/I$ \cite{RRoth2} which is in-line with our calculations. 

We also performed an additional calculation for the large $I=N=50$ lattice using the DMRG-C basis. These results are not shown in the plots because there are no sizable deviations to calculations with the DMRG-B basis. Only for the condensate fraction at $U/J<3$ the DMRG-C basis yields slightly larger values, e.g. $f_c=0.68$ instead $f_c=0.63$ at $U/J=1$. This indicates the slower convergence of the DMRG algorithm in the weakly interacting regime. The results for all other observables remain completely unchanged when going to the larger DMRG-C basis.

\paragraph{Mott-insulator to quasi Bose-glass ($U/J=30$).}
The lower row of Fig.~\ref{DMRGscaling} reveals that the energy gap as well as the maximum number fluctuation do not change with the size of the lattice across the Mott-insulator to quasi Bose-glass transition. The condensate fraction exhibits the previously mentioned $1/I$ scaling which is characteristic for large values of $U/J$. The visibility in the small lattice is again slightly smaller compared to the two larger lattices.

Considering this analysis, we conclude that calculations including $N=30$ particles on $I=30$ lattice sites are sufficient  to describe realistic experiments. Firstly, because this system size is right in the experimental range. And secondly, when going to larger systems, there are only small and predictable changes for the condensate fraction whereas all other observables remain unchanged.

\section{Ab-Initio Phase Diagrams}

After the validation of our framework we now discuss the experiment-specific phase diagram of an ultracold $^{87}$Rb gas with scattering length $a_s=109 \; r_{\text{Bohr}}$ in an optical lattice with wavelength $\lambda_2=800$ nm. The superlattice topology is generated by an additional laser with wavelength $\lambda_1=1000$ nm and relative phase shift of $\phi=\pi/4$. The respective optical  potential depth resulting from the two lasers are given by the dimensionless parameters $s_2$ and $s_1$. The remaining transverse lasers of the optical trap enter via the transverse trapping frequency $\omega_{\perp}$ which is chosen to be $2 \pi \times 17$ kHz. Initially, the longitudinal trapping frequency $\omega_x$ is set to $0$ Hz.

\begin{figure*}
\includegraphics[width=0.9\textwidth]{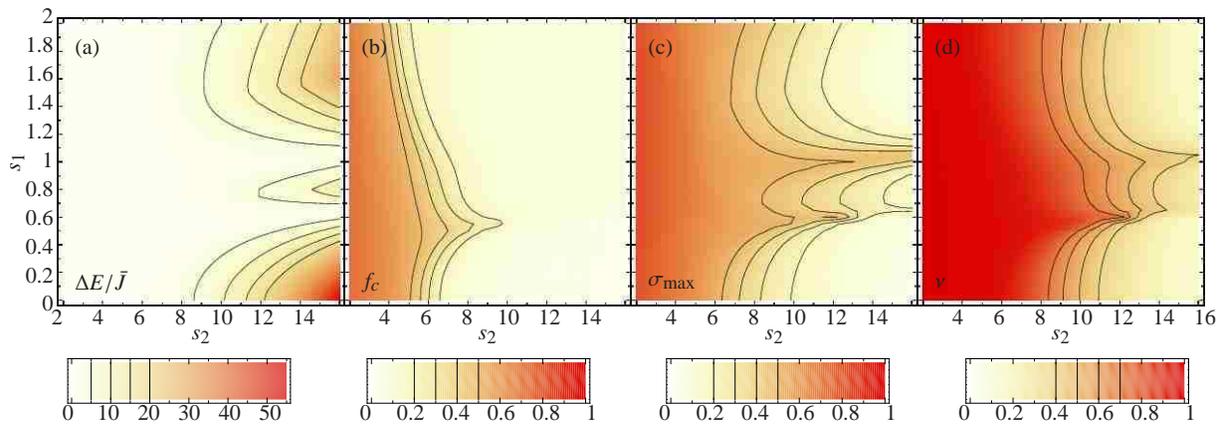}
\caption{\label{PhaseDiag0}(color online) Phase diagram in terms of energy gap $\Delta E/\bar{J}$, condensate fraction $f_c$, maximum number fluctuation $\sigma_{\text{max}}$, and visibility $\nu$ for $I=N=30$, $a_s=109 \; r_{\text{Bohr}}$, $\omega_{\perp}=2 \pi \times 17$ kHz, and  $\omega_x=0$ Hz.}
 \end{figure*}

We have already used these parameters in Fig.~\ref{ComparePic} to compare the experiment-specific phase diagram spanned by $s_2$ and $s_1$ with a generic phase diagram spanned by $U/J$ and $\epsilon_{\text{max}}/J$ neglecting the site dependence of $U$ and $J$. Both panels of Fig.~\ref{ComparePic} show the energy gap $\Delta E$ for $I=N=30$ obtained from a DMRG calculation using the DMRG-B basis. Since the variation of $s_2$ and $s_1$ affects all Hubbard parameters simultaneously, the ($s_2,s_1$) phase diagram is distorted in comparison to the ($U/J,\epsilon_{\text{max}}/J$) phase diagram. However, the ($s_2,s_1$) phase diagram reveals that all relevant quantum phases are accessible through the variation of the intensity of the two longitudinal lasers alone, while keeping the other parameters fixed.

A detailed analysis of the phase diagram for this set of parameters is given in Figs.~\ref{PhaseDiag0}(a)-(d), where we show the energy gap, the condensate fraction, the maximum number fluctuation, and the visibility, respectively.

The superfluid (SF) phase is characterized by a vanishing energy gap, large condensate fraction, large number fluctuations, and maximum visibility. Although we do not compute the most stringent order parameter for the SF phase --- the superfluid fraction \cite{Fisher,RRoth2,Roux} --- the above signatures allow us to identify the SF phase in the region of small $s_2$ up to $s_2\lesssim 6$ in the whole range of $s_1$ shown here. Due to the shallow optical potential in this region the tunneling term in the Hubbard Hamiltonian (\ref{HubbardHamil}) dominates. This results in a coherent many-body state which is a prerequisite for the SF phase.  For $s_2=6$ along $0 < s_1 \leq 2$ the mean interaction energy is $\bar{U}/\bar{J} \approx 4.5$ which explains the presence of the SF phase in the whole range of $s_1$.

In a homogeneous lattice ($s_1=0$ or $\epsilon_{\text{max}}/J=0$) a transition from the SF phase to the homogeneous Mott-insulating (MI) phase occurs around $U/J \approx 5$ \cite{Roux,RRoth2} which corresponds to $s_2=6.25$. This is in-line with our results, because around $s_2\approx 6$ the energy gap steeply increases while the condensate fraction, the number fluctuations, and the visibility decrease. At $s_2=16$ and $s_1=0$ the ratio of $U/J$ is 60 and the system is deep in the homogeneous MI phase showing the characteristic large energy gap and vanishing number fluctuations, condensate fraction, and visibility.

If we now increase $s_1$ at fixed $s_2=16$, the modulation of the site-dependent Hubbard parameters grows rapidly and at $s_1 \approx 0.6$ the spread of the on-site energies becomes comparable to the average interaction energy, i.e., $\epsilon_{\text{max}} / \bar{J} \approx \bar{U} / \bar{J}$. Thus, despite the strong repulsive interaction, it becomes advantageous to move an atom from a site with large on-site energy to an already occupied site with small on-site energy. Due to this redistribution of particles the homogeneous MI phase is broken up and the transition to the quasi Bose-glass (BG) phase occurs. The commensurate superlattice defined by $\lambda_2=800$ nm, $\lambda_1=1000$ nm, and $\phi=\pi/4$ exhibits only 5 different on-site energies. This small set of on-site energies leads to extended domains in the phase diagram. Two of these domains are visible in Fig.~\ref{PhaseDiag0}(a). Only in the transition region between them the energy gap vanishes.

The genuine Bose-glass phase occurs only in an infinite lattice with random on-site energies. It is marked by a completely vanishing energy gap. Intuitively this results from a continuous distribution of on-site energies permitting the construction of excited states by moving particles to sites with infinitesimally larger on-site energies associated with infinitesimally small excitation energies. We have approached this limit using an incommensurate superlattice in a previous publication \cite{Schmitt}.

\section{Longitudinal Trapping Frequency $\omega_x$}

\begin{table}
\begin{tabular}{|l||c|c|c|c|}
\hline
$s_2$					&	$2$	&	$10$	&	$12$	&	$16$	\\
\hline\hline
$\bar{J}/E_{r_2}$			&	$0.1428$&	$0.0192$&	$0.0123$&	$0.0053$\\
\hline
$\bar{U}/E_{r_2}$			&	$0.1624$&	$0.2505$&	$0.2966$&	$0.3232$\\
\hline
$\bar{U}/\bar{J}$			&	$1.1378$&	$14.623$&	$24.222$&	$60.636$\\
\hline
\hline
\multicolumn{5}{|c|}{$\omega_x=2 \pi \times 25$ Hz}\\
\hline
$\epsilon_{\text{max}}/E_{r_2}$	&	$0.02431$&	$0.02431$&	$0.02431$&	$0.02431$\\
\hline
$\epsilon_{\text{max}}/\bar{J}$	&	$0.17052$&	$1.26739$&	$1.98446$&	$4.55600$\\
\hline
\hline
\multicolumn{5}{|c|}{$\omega_x=2 \pi \times 50$ Hz}\\
\hline
$\epsilon_{\text{max}}/E_{r_2}$	&	$0.09737$&	$0.09737$&	$0.09737$&	$0.09737$\\
\hline
$\epsilon_{\text{max}}/\bar{J}$	&	$0.68196$&	$5.06927$&	$7.937412$&	$18.2389$\\
\hline
\hline
\multicolumn{5}{|c|}{$\omega_x=2 \pi \times 75$ Hz}\\
\hline
$\epsilon_{\text{max}}/E_{r_2}$	&	$0.21909$&	$0.21909$&	$0.21909$&	$0.21909$\\
\hline
$\epsilon_{\text{max}}/\bar{J}$	&	$1.53407$&	$11.4048$&	$17.8576$&	$41.0344$\\
\hline
\hline
\multicolumn{5}{|c|}{$\omega_x=2 \pi \times 100$ Hz}\\
\hline
$\epsilon_{\text{max}}/E_{r_2}$	&	$0.38950$&	$0.38950$&	$0.38950$&	$0.38950$\\
\hline
$\epsilon_{\text{max}}/\bar{J}$	&	$2.72634$&	$20.27242$&	$31.7430$&	$72.9423$\\
\hline
\end{tabular}
\caption{\label{TabOmegaX}Comparison of the Hubbard parameters to analyze effect of the longitudinal trapping potential. The parameters are: $\lambda_2=800$ nm, $\lambda_1=1000$ nm, $\phi=\pi/4$, $s_1=0$, $\omega_{\perp}=2\pi \times 17$ kHz, and mass and scattering length of $^{87}$Rb.}
\end{table}

Since the aim of this manuscript is the calculation of an experiment specific phase diagram for a realistic experimental setup, it is compulsory to consider an additional magnetic trapping potential and the intensity variation of the optical lattice through the focusing of the laser beams. To this end we have introduced a harmonic potential with frequency $\omega_x$ in Eq.~(\ref{OptPot}). Typical experimental parameters range from $\omega_x=2 \pi \times 10$ Hz to $2\pi \times 75$ Hz \cite{Stoeferle,Lye,Fallani}.

To get a impression of the energy scales, we show some values for the Hubbard parameters obtained by our band structure approach in Tab. \ref{TabOmegaX}. By setting $s_1=0$ the on-site energies are solely due to the additional harmonic potential. At the outer rims of the lattice (sites $1$ and $30$) they have the value $\epsilon_{\text{max}}$.

\begin{figure*}
\includegraphics[height=0.709\textheight]{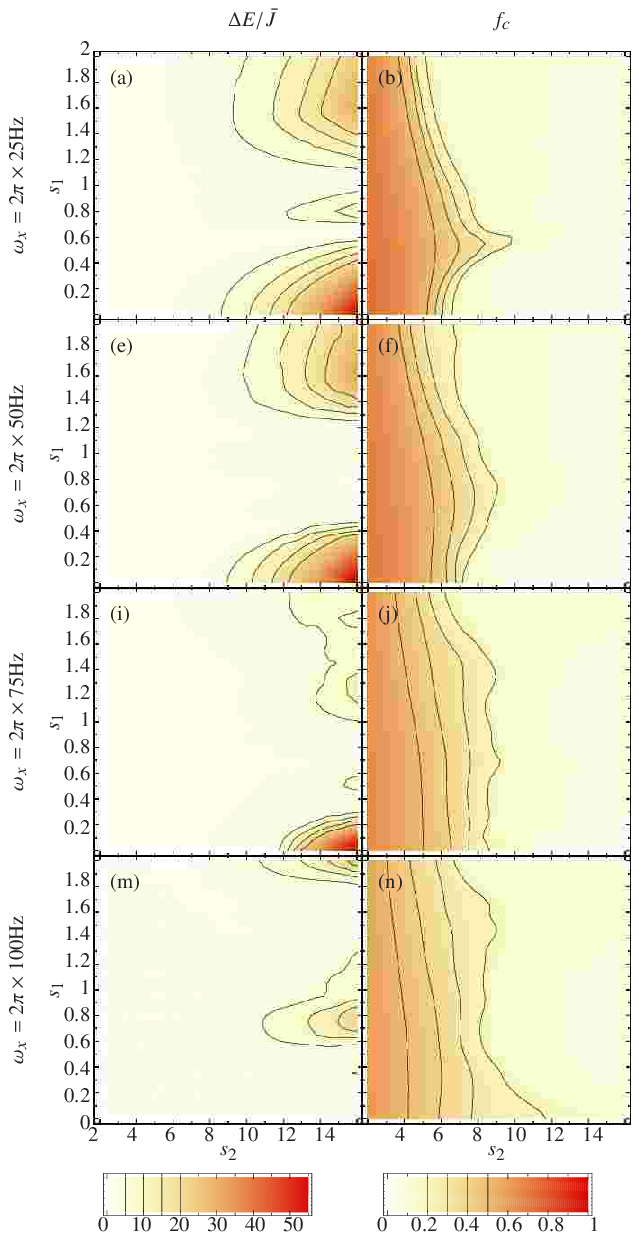}\hspace{-3pt}%
\raisebox{1.5pt}{\includegraphics[height=0.7\textheight]{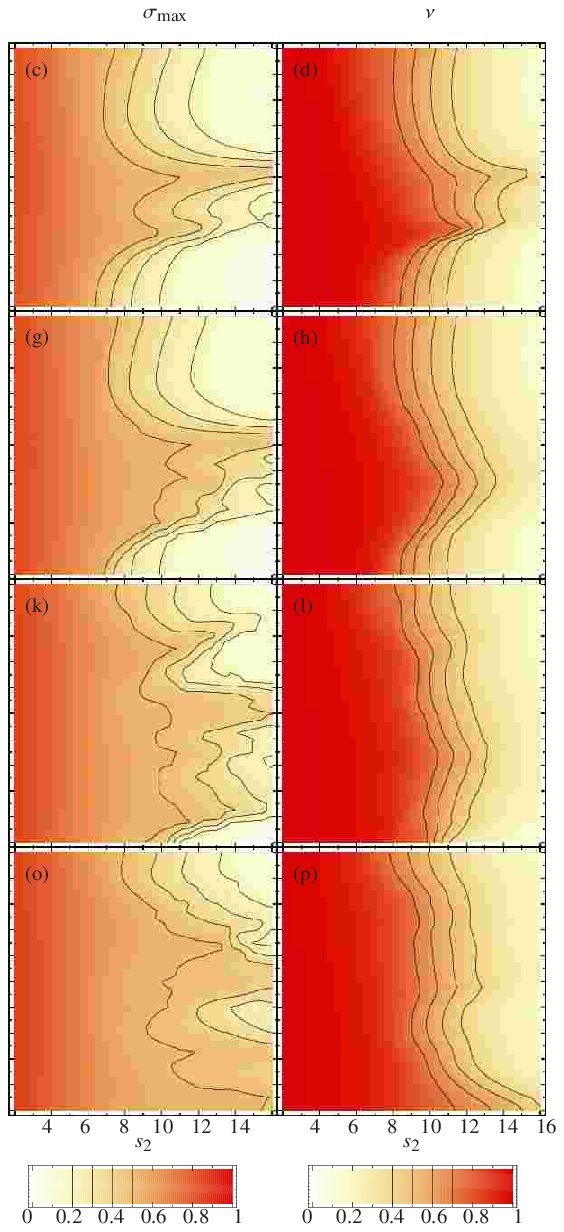}}
\caption{\label{PhaseDiag100}(color online) Phase diagram in terms of energy gap $\Delta E/\bar{J}$, condensate fraction $f_c$, maximum number fluctuation $\sigma_{\text{max}}$, and visibility $\nu$ for $I=N=30$, $a_s=109 \; r_{\text{Bohr}}$, $\omega_{\perp}=2 \pi \times 17$ kHz. First row: $\omega_x= 2 \pi \times 25$ Hz, second row: $\omega_x= 2 \pi \times 50$ Hz, third row: $\omega_x= 2 \pi \times 75$ Hz, fourth row: $\omega_x= 2 \pi \times 100$ Hz.}
\end{figure*}

Up to $\omega_x=2 \pi \times 25$ Hz, $\epsilon_{\text{max}} / \bar{J}$ is an order of magnitude smaller than $\bar{U} / \bar{J}$. For this reason, the phase diagram remains practically unaltered between $\omega_x= 0$ Hz and $2 \pi \times 25$ Hz as can be seen by comparing Figs.~\ref{PhaseDiag0}(a)-(d) and \ref{PhaseDiag100}(a)-(d).
%
For $\omega_x=2 \pi \times 50$ Hz Tab. \ref{TabOmegaX} shows $\epsilon_{\text{max}} / \bar{J}$ is still about a factor 3 smaller than $\bar{U} / \bar{J}$. As a consequence the onset of the BG phase in Figs.~\ref{PhaseDiag100}(e)-(h) already appears at $s_1\approx 0.4$ instead of $s_1\approx 0.6$ for $\omega_x=0$ Hz. Besides the earlier onset of the BG phase also its gross structure changes. The lobe around $s_1= 0.6$ in Fig.~\ref{PhaseDiag100}(e) is suppressed compared to the calculations for $\omega_x<2 \pi \times 50$ Hz . Also the maximum fluctuations indicate that the redistribution of particles becomes smoother.  This is because for $\omega_x=0$ Hz the superlattice topology exhibits only 5 different on-site energies. With the additional harmonic potential the number of different on-site energies increases and, therefore, the extended domains in the BG phase shrink.
%
For $\omega_x = 2 \pi \times 75$ Hz  the parameters $\epsilon_{\text{max}} / \bar{J}$ and $\bar{U} / \bar{J}$ become comparable and the phase diagram changes dramatically. In Fig.~\ref{PhaseDiag100}(i) the homogeneous MI domain shrinks to a small region ($s_2=12-16$ and $s_1=0-0.2$). Furthermore, a clear detection of the BG phase becomes difficult since the characteristic increase of the visibility along the MI to BG transition is no longer visible in Fig.~\ref{PhaseDiag100}(l).
%
Table \ref{TabOmegaX} reveals that at $\omega_x =2 \pi \times 100$ Hz the on-site energies $\epsilon_{\text{max}}$ clearly dominate the energy scale. Thus, even the transition from the SF phase to the MI phase is no longer observable in the investigated parameter range of $s_2$. The MI domain has completely disappeared in Fig.~\ref{PhaseDiag100}(m) and the visibility remains large throughout the whole range of $s_2$ for $s_1<0.2$.

From the above discussion we conclude that the smaller the longitudinal trapping frequency $\omega_x$, the easier is a clear distinction between the SF, MI and BG phases. Thus, any experiment with a focus on the phase diagram of ultracold atoms in an optical superlattice should be designed such that the longitudinal trapping frequency is kept small.

\section{Transverse Trapping Frequency $\omega_{\perp}$}

\begin{figure*}
\includegraphics[width=0.9\textwidth]{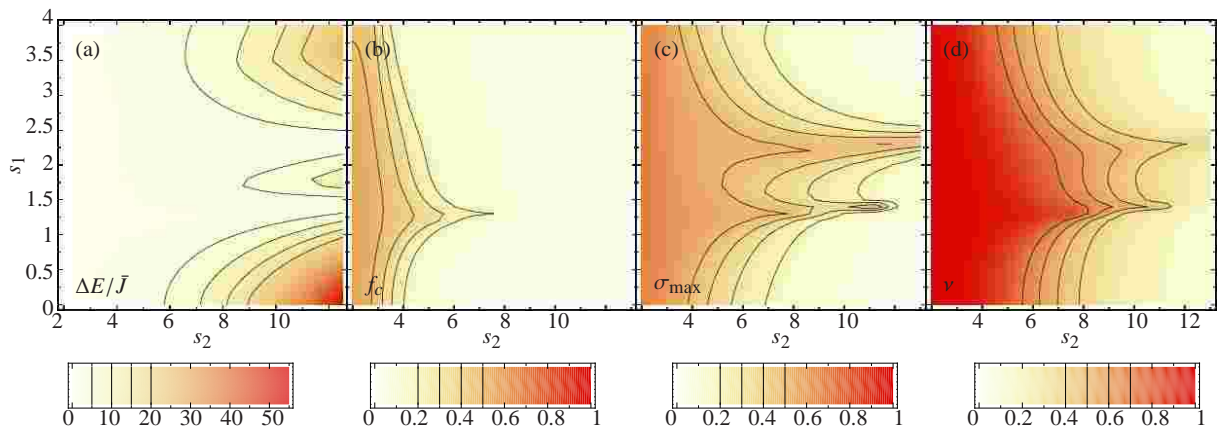}
\caption{\label{PhaseDiagOmegaP}(color online) Same parameter set as in Fig.~\ref{PhaseDiag0} but with transverse trapping frequency $\omega_{\perp}= 2 \pi \times 40$ kHz. While the gross structure of the phase diagram is independent of $\omega_{\perp}$, the scales of the $s_2$ and $s_1$ axes change.}
\end{figure*}

Finally we study the dependence of the $(s_1,s_2)$ phase diagram on the intensity of the transverse lasers through a variation of the transverse trapping frequency $\omega_{\perp}$. For the sake of simplicity we assume $\omega_x=0$ Hz.

From Eq.~(\ref{hubbardparam}) it follows that the interaction energy $U_i$ is proportional to $\omega_{\perp}$ while $\epsilon_i$ and $J_i$ are independent of  $\omega_{\perp}$. A larger value of $\omega_{\perp}$ will, therefore, shift the SF to MI transition towards smaller $s_2$. Also the spread of the on-site energies must increase to overcome energy cost of a double occupancy and consequently the MI to quasi BG transition will shift towards larger $s_1$. In Fig.~\ref{PhaseDiagOmegaP} we show the phase diagrams for $\omega_{\perp}=2 \pi \times 40$ kHz and all other parameters unchanged. The gross structure of the phase diagram remains the same. However, in accordance with our considerations, the energy gap is $\Delta E/ \bar{J}=60$ already at $s_2=12.5$ instead of $s_2=16$ for $\omega_{\perp}=2 \pi \times 17$ kHz (both $s_1=0$). Furthermore, the transition from the homogeneous MI to the quasi BG phase occurs around $s_1 \approx 1.5$ as compared to $s_2=0.6$ for  $\omega_{\perp}=2 \pi \times 17$ kHz.

As a consequence of the dependence of the ($s_2,s_1$) phase diagram on the transverse trapping frequency, the precise identification of the phase boundaries is intimately connected to a well defined value of $\omega_{\perp}$.

\section{Summary \& Conclusions}

We have studied the experiment-specific phase diagram of ultracold $^{87}$Rb atoms in an one-dimensional two-color superlattice with respect to the parameters of the experiment. Band structure calculations were employed to obtain the generic parameters of the Hubbard model from the experiment-specific parameters. These band structure calculations were also used to confirm the applicability of the Hubbard model in the investigated parameter range.

In order to solve the many-body problem for realistic lattice lengths and particle numbers we have used the density-matrix renormalization-group algorithm. Through a thorough benchmark of our DMRG calculations we demonstrated that all observables are perfectly converged and can practically be considered as exact solutions of the many-body problem. Furthermore, a detailed finite-size analysis  for all observables has underlined the significance of our results for realistic experimental system sizes.

Our calculations of the phase diagrams show that all relevant quantum phases can be addressed by only varying the intensities of the two lasers that generate the optical superlattice. For a longitudinal trapping frequency $\omega_x<25$ Hz all different quantum phases can be clearly distinguished by means of the presented observables. However, larger values of the longitudinal trapping frequency lead to radical changes in the structure of the phase diagram and make a clear identification of the Bose-glass phase impossible.

We also showed that the gross structure of the phase diagram does not depend on the transverse trapping frequency $\omega_{\perp}$, i.e., the intensity of the lasers in the directions perpendicular to the 1D lattice. However, due to the linear dependence of the interaction energy on the transverse trapping frequency, the position of the transition lines in the phase diagram crucially depend on that parameter.

\section*{Acknowledgment}

Supported in part by the ExtreMe Matter Institute EMMI in the framework of the Helmholtz Alliance HA216/EMMI and the DFG Sonderforschungsbereich 634.



\end{document}